\PassOptionsToPackage{final}{graphicx}
\documentclass[sigconf,natbib=false,authorversion,nonacm]{acmart}

\usepackage[backend=biber,bibencoding=utf8,style=ACM-Reference-Format,natbib,mincitenames=1,maxcitenames=2,maxbibnames=3,autocite=inline,sortcites,labelnumber]{biblatex}
\DeclareFieldFormat*{citetitle}{\emph{#1}}

\hypersetup{draft=false}

\usepackage[obeyDraft,colorinlistoftodos,prependcaption]{todonotes}

\usepackage[group-separator={,},group-minimum-digits={3}]{siunitx}
\usepackage{nicefrac}
\usepackage{textcomp}
\usepackage{soul}
\usepackage{mathrsfs}
\usepackage{amsmath}
\usepackage{amsfonts}
\usepackage{mathtools}
\usepackage{numname}
\usepackage[capitalize]{cleveref}

\crefname{listing}{listing}{listings}
\Crefname{listing}{Listing}{Listings}
\crefname{sublisting}{listing}{listings}
\Crefname{sublisting}{Listing}{Listings}

\usepackage{comment}
\usepackage{ellipsis}

\usepackage{algorithm}
\usepackage{algorithmic}

\crefname{ALC@unique}{line}{lines}
\Crefname{ALC@unique}{Line}{Lines}

\crefname{paragraph}{section}{sections}
\Crefname{paragraph}{Section}{Sections}

\graphicspath{{./graphics/}} %

\usepackage[shortlabels,inline]{enumitem}

\usepackage[font=small,skip=0.75em]{caption}
\crefname{section}{\S}{\S}
\crefformat{section}{\S#2#1#3}
\Crefname{section}{\S}{\S}
\Crefformat{section}{\S#2#1#3}
\usepackage[compact,uppercase]{titlesec}

\newcommand{\ti}[1]{\texttt{#1}}
\newcommand{\pyi}[1]{\texttt{#1}}

\usepackage{booktabs}
\usepackage{tabulary}
\usepackage{threeparttable}
\usepackage{multirow}

\newcommand{\GH}{\href{http://github.com}{GitHub}}
\newcommand{\tool}{\texttt{QuerTCI}}
\newcommand{\YouTube}{\url{https://youtu.be/fADKSxn0QUk}}

\graphicspath{ {./figures/} }

\setcopyright{acmcopyright}
\copyrightyear{2022}
\acmYear{2022}

\begin{document}

\title{QuerTCI\@: A Tool Integrating GitHub Issue Querying with Comment Classification}

\author{Ye Paing}
\affiliation{%
  \institution{CUNY Hunter College}
  \city{New York}
  \state{NY}
  \country{USA}
}
\email{Ye.Paing89@myhunter.cuny.edu}

\author{Tatiana Castro V\'{e}lez}
\affiliation{%
  \institution{CUNY Graduate Center}
  \city{New York}
  \state{NY}
  \country{USA}
}
\email{tcastrovelez@gradcenter.cuny.edu}

\author{Raffi Khatchadourian}
\orcid{0000-0002-7930-0182} %
\affiliation{%
  \institution{CUNY Hunter College}
  \city{New York}
  \state{NY}
  \country{USA}
}
\email{raffi.khatchadourian@hunter.cuny.edu}

\begin{abstract}

    Empirical Software Engineering (ESE) researchers study (open-source) project issues and the comments and threads within to discover---among others---challenges developers face when incorporating new technologies, platforms, and programming language constructs. However, such threads accumulate, becoming unwieldy and hindering any insight researchers may gain.
While existing approaches alleviate this burden by classifying issue thread comments, there is a gap between searching popular open-source software repositories (e.g., those on \GH) for issues containing particular keywords and feeding the results into a classification model. This paper demonstrates a research infrastructure tool called \tool\ that bridges this gap by integrating the \GH\ issue comment search API with the classification models found in existing approaches. Using queries, ESE researchers can retrieve \GH\ issues containing particular keywords, e.g., those related to a specific programming language construct, and, subsequently, classify the discussions occurring in those issues. We hope that ESE researchers can use our tool to uncover challenges related to particular technologies using specific keywords through popular open-source repositories more seamlessly than previously possible. A tool demonstration video may be found at: \YouTube.

\end{abstract}



\maketitle

\section{Introduction}

Issue tracking systems, e.g., \GH\ issues, enable the discussion of software problems. Empirical Software Engineering (ESE) researchers have engaged in several activities related to mining software repositories (MSR), including studying (open-source) project issues. Researchers examine comments and threads contained in issues to discover the challenges developers face in writing software. For example, developers may struggle with incorporating new technologies, platforms, and programming language constructs and document and discuss their progress in issue ``tickets.''

Unfortunately, such issue discussion threads accumulate over time and thus can become unwieldy, hindering insights researchers may gain from them. Moreover, issues can be unlabeled or improperly named, making it difficult to understand the problem at hand.
Approaches (e.g.,~\cite{Arya2019}) exist to alleviate this burden by classifying issue thread comments; however, there is a gap between searching popular open-source software repositories (e.g., those on \GH) for issues containing particular keywords and feeding the results into a classification model. In this paper, we demonstrate \tool, a \textbf{Quer}y-based \textbf{T}ool for \textbf{C}lassifying \GH\ \textbf{I}ssue thread comments that bridges this gap by integrating the \GH\ issue comment search API with a classification model. While the default classification model used by \tool\ is the one developed by~\citet{Arya2019}, it can also use other issue comment classification models. \tool\ is a Python-based tool that queries \GH's search API for issues and comments relating to a query string that the user provides. Then, it automatically preprocesses (i.e., parses, cleans, tokenizes) each line of issue comments retrieved from the \GH\ API and runs them through the pre-trained NLP model for classification.

\begin{figure}[t]
    \centering
    \includegraphics[width=0.75\linewidth]{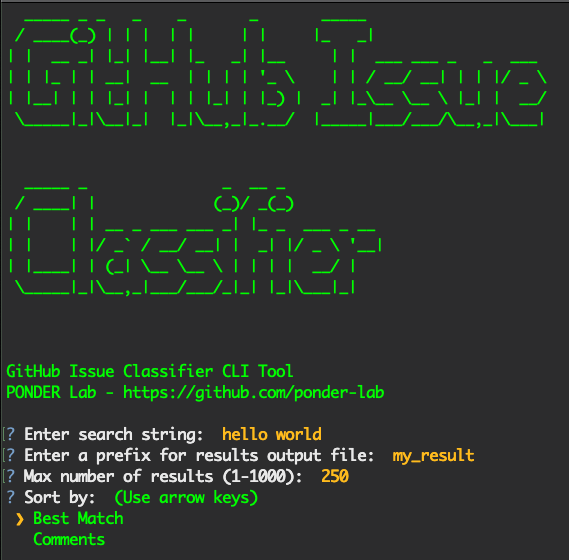}
    \caption{\tool\ interactive command-line interface.}\label{fig:cli}
    \Description{\tool\ interactive command-line interface.}
\end{figure}

\tool\ is highly-customizable---supporting a wide range of additional functionalities---and works in either interactive and batch (non-interactive) modes. As shown in \cref{fig:cli}, users can limit number of issues retrieved (the \GH\ API supports up to \num{1000}), as well as change the sorting criteria (which is important in capped queries like \GH). Users may also omit particular issue comment classification categories from the results, e.g., retrieving issues that have at least one comment corresponding to a ``solution discussion''~\cite{Arya2019}.

Using queries, \tool\ enables ESE researchers to retrieve \GH\ issues containing particular keywords, e.g., those related to a certain programming language construct, and subsequently classify the kinds of discussions occurring in those issues in an integrated manor. Our hope is that, by using \tool\ as part of a broader research infrastructure, ESE researchers can uncover challenges related to particular technologies using certain keywords through popular open-source repositories more seamlessly than previously possible. It alleviates the required leg work of data querying and preparation in order for the data to be:
\begin{enumerate*}[(i)]
    \item compatible with the underlying comment classification model and
    \item related to particular programming language constructs (ala \citetitle{Casalnuovo2017}~\cite{Casalnuovo2017}).
\end{enumerate*}
\tool\ is open-source and publicly available~\cite{Paing2021}, and a demonstration video may be found at: \YouTube.

\section{Envisioned Users}\label{sec:users}

Since we foresee \tool\ being part of a broader research infrastructure, our envisioned users are mainly ESE researchers seeking to unearth challenges developers face in particular situations or using specific technologies. For example, ESE researchers may be interested in discovering---using a keyword-based search---the kinds of discussions surrounding a particular Application Programmer's Interface (API), programming language feature, or new platform version. Using \tool, they are able to receive---using NLP to automatically identify the topics/semantics of what is being discussed---a ``quick gist'' of otherwise long discussion threads commonly found in \GH\ issues. ESE researchers---on a relatively large scale---can then quantify categories of discussions taking place under \GH\ issues containing keywords of interest.

To further understand developer challenges, ESE researchers may filter for issues containing keywords corresponding to certain language constructs and having particular comment classifications. Then, they may use manual inspection to further investigate. \tool\ empowers ESE researchers to narrow the scope of manual investigation so that they may focus on \GH\ issues with the most relevant discussions. For instance, we may be interested in \GH\ issues involving Java 8 streams (e.g., \texttt{stream()}, \texttt{parallelStream()}, \texttt{Collectors}) that include no solution discussion. The resulting set can then be further manually inspected by users to understand why issues involving these constructs cannot be solved.

Another class of users may be practicing Software Engineers that are tackling common problems pertaining to certain topics. Using \tool, they can summarize issues that may arise for a particular topic/query string of interest. Furthermore, Software Engineers can use \tool\ to filter out certain discussion types (e.g., ``Social Discussion''~\cite{Arya2019}) from the tool's output, thereby saving time that might have been spent combing through large discussion threads.

Lastly, \tool\ may prove useful to (CS) students studying a particular topic---looking to platforms, e.g., \GH, to gain insight into what experienced Software Engineers are discussing. Using our tool, students can focus on issue threads that pertain to relevant categories (e.g., ``Solution Discussion,'' ``Bug Reproduction''~\cite{Arya2019}). Such expert developer discussion may prove useful to students learning a new programming language, struggling with using a new framework or library (APIs), or adopting a new software platform.

\section{Addressed Empirical Software Engineering Research Challenges}\label{sec:motive}

With \tool, we aim to reduce the time spent by researchers---and perhaps Software Engineers---in combing through long issue discussion threads for answers relating to their topic of interests. \tool\ was created to help supplement ESE researchers in quantifying and ultimately understanding long issue discussion threads within \GH\ issues containing particular keywords (e.g, related to certain APIs). A challenge that ESE researchers face is the time it takes to fully digest and understand issue discussion threads using manual inspection. %
Although \GH\ issues may be ``tagged,'' such tags may either be inaccurate nor reflect how issue discussions evolves. Moreover, issue tags, e.g., ``bug,'' ``enhancement,'' relate to the \GH\ issue as a whole and thus may not represent the \emph{conversations} occurring within the issue. As (manual) empirical studies are typically labor and resource intensive, ESE researchers can use \tool\ to reduce the search space needed to discover answers to their (research) questions. Researchers can thus query \GH\ issues for keywords and subsequently rely on a pre-trained NLP model for issue thread discussion categorization. Doing so can limit the search space (i.e., issues and discussion threads) necessary for ESE researchers to (manually) inspect. %

For example, suppose that we are interested in challenges data scientists face in using a particular TensorFlow API, e.g., \texttt{tf.module}. Simply using \tool\ to query for the string ``tf.module'' would yield a large amount of \GH\ issue threads that would automatically be categorized using a pre-trained NLP model, allowing researchers to save time, gain an overview of the nature of the issues surrounding this particular query, and choose a proper subset of \GH\ issues to (manually) examine in further detail.

\section{Implementation}\label{sec:impl} %

\begin{figure}[t]
    \centering
    \includegraphics[width=1.0\linewidth]{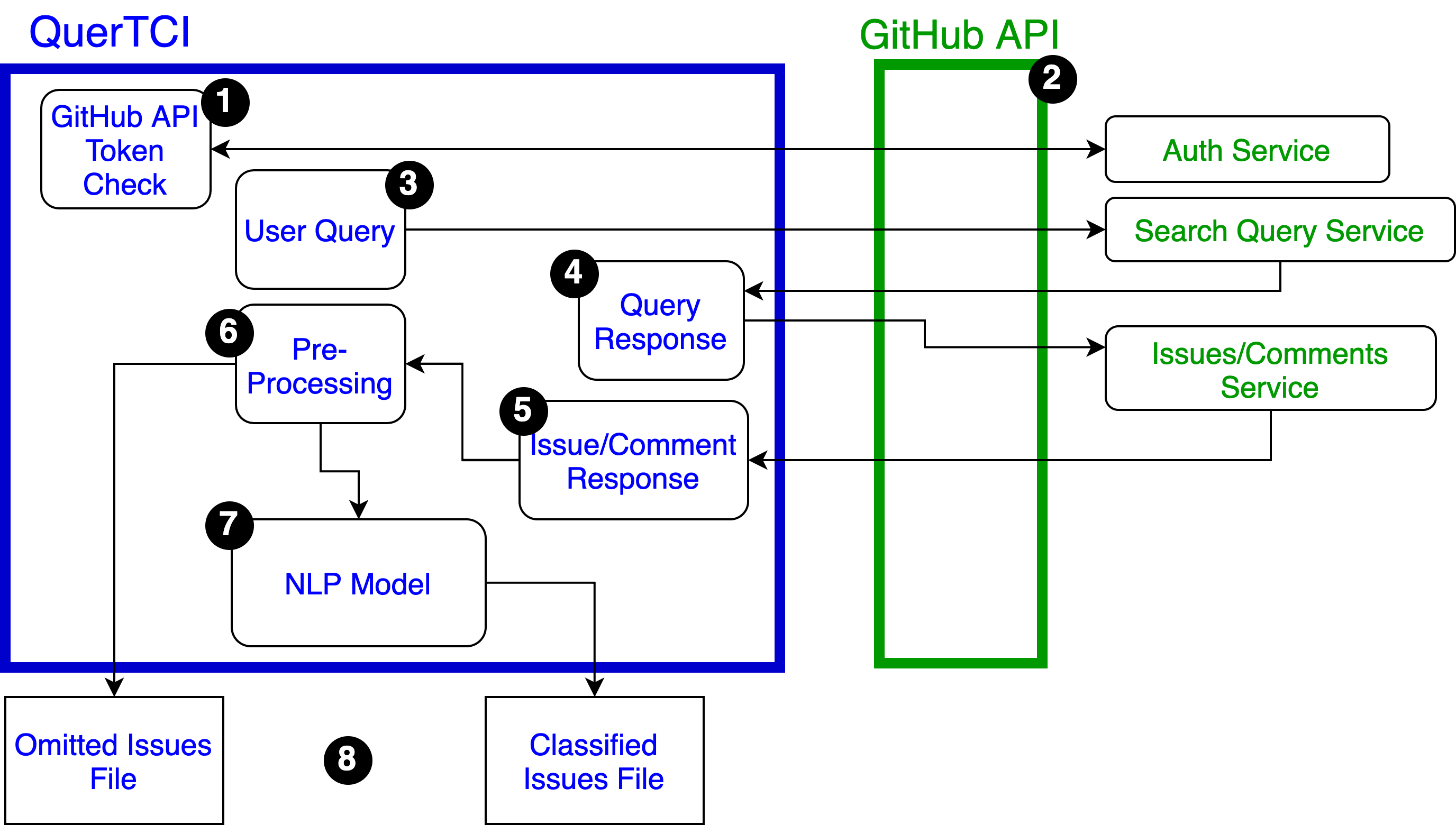}
    \caption{\tool\ high-level architecture.}\label{fig:arch}
    \Description{\tool\ high-level architecture.}
\end{figure}

\Cref{fig:arch} depicts the overall architecture of \tool, which includes querying against \GH's query service API to search for results pertaining to the input query string, preprocessing the retrieved data locally, and running the retrieved data through a pre-trained NLP model for classification. Finally, \tool\ writes the results to output files on the local file system.

\textit{NLP Model, Constraints, Serialization \& CLI Prototyping.}\label{sec:nlp}
The \tool\ implementation uses a pre-built, customizable NLP model to classify issue comments. By default, the tool integrates a model provided by \citet{Arya2019}, but users may provide their own. The models are serialized using the \ti{sklearn}~\cite{SLD2022} and \ti{pickle}~\cite{PSF2022} Python libraries---persisting them into a file that is ``imported'' into \tool. The pre-built (input) model should only be responsible for classification; models should not employ any preprocessing or tokenization of the input strings (i.e., those representing individual lines of \GH\ issue comment threads). Preprocessing and tokenization---detailed shortly---are done by \tool\ before running them through the model. This implementation step involved thorough testing to ensure that integrating with the imported model did not generate any errors or data corruption. \tool's CLI was prototyped using an iterative process. To assess usability, we surveyed several ESE researchers in our lab (independent of this project) to understand the options needed and ease-of-use.

\textit{Interfacing with the GitHub API.}
Our tool interfaces with the \GH\ API (\cref{fig:arch}, step 2).
Our implementation lifts several burdens of ESE researchers seeking to programmatically query \GH, including API query
throttling, API key transfer, and HTTP request authentication (\cref{fig:arch}, step 1).
Then, a query encompassing a search term (\cref{fig:arch}, step 3) is used to retrieve a list of \GH\ issue threads (e.g., pull/patch requests,\footnote{\GH\ treats issues and pull requests similarly.} issue discussion), where comments for each of the returned issue threads are extracted. This step is accomplished via several REST API endpoints, e.g., \ti{/issues} that, given a query parameter \ti{q}, returns a list of issues that are related to the desired query parameter string~\cite{GitHub2021}.

With the results from the previous search query (\cref{fig:arch}, step 4), we then query the \ti{/comments} REST API endpoint for each of the retrieved issues. This endpoint is provided as part of the issue results. Querying this endpoint returns the list of comments for each of the issues (\cref{fig:arch}, step 5), thus allowing us to extract comment strings for further processing, cleaning, and tokenizing before they are run through the classification model. At this step, we also filter out noisy comments, e.g., for query strings that include punctuation,\footnote{\GH\ ignores punctuation in all query strings.} as well as issues that do not contain any discussion.

\textit{Data Preprocessing \& Tokenization.}
With the list of comments previously retrieved from the \ti{/comments} REST API call, we then preprocess and clean the returned data, removing any noise and unnecessary stop words (\cref{fig:arch}, step 6). We use the list of stop words included in the \ti{NLTK} library~\cite{Loper2002}, augmenting it with several of our own custom words to further help reduce noise within the comment corpus. Additionally, we tokenize certain (common) strings in
order to extract the essence of the \GH\ issue text. This process mainly centers around tokenizing screen (\GH\ user) names,
URLs,
quotes
(both single and double) and code snippets (strings beginning with back ticks). Each token is then replaced with token names, e.g., \ti{USER\_NAME}, \ti{URL}, \ti{QUOTE}, \ti{CODE}.
Lastly, due to the way \GH\ processes queries---using relaxed matching---\tool\ further filters out issue comments that do not contain the original query string. This is particularly important for queries representing programming language constructs or API calls, which typically include punctuation.
Issues not matching this stricter check are omitted and stored in a corresponding ``omitted'' file for users to inspect further if necessary (\cref{fig:arch}, step 8).

\textit{Model Classification \& Result Output.}\label{sec:res}
\begin{table*}
    \centering
    \caption{Example result CSV file snippet. Issue URLs not shown. Column \textbf{id} is the GitHub issue identifier.}\label{tab:res}
    \footnotesize
    \begin{tabular}{@{}lp{4.9in}l@{}}
	\toprule
	id & comment line & category \\ \midrule
	415902593 & however get u step closer running original code actual error message tensorboard propagate ui CODE & Observed Bug Behavior \\
	415902593 & i think simplest fix around would call trace\_on trace\_export separately around graph call so something like & Workarounds \\
	417390174 & some detail i using subclassed model complex valued data & Motivation \\
	740456602 & removing tf.function decorator viable workaround best practice a related issue URL tensorflow issues/27120 & Potential New Issues \& Requests \\
	755665148 & 74 fix acquisition optimizer & Solution Discussion \\
	767685452 & SCREEN\_NAME still issue latest version coremltools if still issue please share additional code show \ldots & Action on Issue \\
	873531279 & i similar issue please help would great & Contribution \& Commitment \\
	947976601 & situation actually much worse i realised CODE CODE the following test pass & Solution Discussion \\
	947976601 & \ldots\ CODE raised even though value execute error branch perhaps due tracing covering every branch this suggests \ldots & Usage \\
	1004824336 & SCREEN\_NAME could specify tensorflow version do use docker & Solution Discussion \\
	1004824336 & tf version CODE unfortunately i use docker & Usage \\
	1004824336 & thx i guess could something wrong pretraining cobblestone because i tested running pre trained cobblestone agent \ldots & Usage \\
	1004824336 & hi i've checked sliced\_trajectory data part correct may i ask chain used pretraining training part forger most likely one? CODE & Usage \\
	1004824336 & SCREEN\_NAME I can reproduce reported behavior docker version also i tried reproduce without docker got error \ldots & Bug Reproduction \\
	1004824336 & yes CODE trajectory i see problem chain for example agent place additional crafting table creating stone pickaxe look first \ldots & Expected Behavior \\
	\bottomrule
    \end{tabular}
\end{table*}
The retrieved issue comment data---now cleaned and preprocessed---is fed it into the model for classification (\cref{fig:arch}, step 7). Classification results are then written to a CSV file. As comments bodies may be lengthy, each comment \emph{line} is classified. We also list the source issue identifiers, including browser- and API-friendly URLs (not shown in \cref{tab:res}) to help easily navigate to the \GH\ issue via a browser for further (manual) inspection.

\Cref{tab:res} portrays an example result CSV file snippet produced by \tool; the complete example file may be found in our dataset~\cite{Paing2022}. Column \textbf{id} represents the unique \GH\ issue identifier, column \textbf{comment line} the preprocessed, tokenized comment text, and column \textbf{category} the classification category as produced using \citet{Arya2019}'s pre-built model. The results were obtained for an empirical study on the challenges facing developers in improving the run-time performance of imperative Deep Learning (DL) code using hybridization~\cite{Velez2022}. The query ``\pyi{tf.function}'' was used (period included) to uncover unsolved \GH\ issues mentioning the \citetitle{Abadi2016}~\cite{Abadi2016} hybridization API keyword. The issues (filtering by \tool) were then manually inspected.

\section{Evaluation}\label{sec:eval}

As \tool\ is in early development stages, a thorough evaluation is pending. However, \tool\ integrates several successful technologies and approaches. The \GH\ API is widely used, both for industry and research. ESE researchers have successfully used the \GH\ API at scale, e.g., \citet{Dilhara2022} use it to discover \num{1000} top-rated Machine Learning (ML) systems comprising \num{58} million source lines of code (SLOC). Furthermore, through qualitative content analysis of \num{15} complex issue threads across three \GH\ projects, \citeauthor{Arya2019}---our default NLP model---uncovered \num{16} different comment classification types, creating a labeled corpus containing \num{4656} sentences~\cite{Arya2019}. Their model has an F-score of \num{0.61} and \num{0.42} for existing and new \GH\ issues, respectively. As mentioned in \cref{sec:res}, our tool has been used in a prior empirical study.

Our isolated preliminary assessment of \tool\ involved a double-blind open card sort between two authors to independently evaluate our tool's integration with \GH\ and the model of \citet{Arya2019} and subsequently assess its accuracy. The authors chose a random selection of issue comment threads based on the same query and independently categorized them. The results were then compared to reach an agreed manual classification. We then used \tool\ to classify the issues comments and compare if \tool\ had categorized these issues in a similar way.

While initial results are promising, we plan to expand the evaluation by involving external ESE researchers. Specifically, we will recruit independent ESE researchers to use \tool\ for an empirical study, e.g., one studying particular API usage. Then, we will recruit other independent ESE researchers \emph{not} using \tool\ as a control. To reduce the number of variables, each of the research teams would perform the \emph{same} study \emph{with} and \emph{without} our tool. However, achieving this goal is highly unlikely as the studies will not be novel. More practically, we will use a mass survey among ESE researchers that have not used our tool and then compare the results with those where the researchers \emph{did} use it. Although the comparison will not be completely isolated, if the scale is large enough, we foresee that the results will nevertheless be useful.

\section{Related Work}\label{sec:related}

\citet{Casalnuovo2017} present \citetitle{Casalnuovo2017}, a tool for processing and classifying \GH\ \emph{commits}. Our tool is for processing and classifying \GH\ issue \emph{comments}. Like our tool, their tool is also motivated by analyzing programming language constructs using (e.g., API) keywords. However, \citetitle{Casalnuovo2017} does not analyze \GH\ SE artifacts at scale; each project repository must be downloaded locally and subsequently (serially) analyzed. \tool, on the other hand, leverages the (indexed) \GH\ API \emph{online}, nearly instantly obtaining \GH\ issues data from thousands of \GH\ projects.

\citet{Arya2019} users must sanitize and manually enter the issue comments as input. Our keyword \emph{query}-based approach \emph{automatically} interfaces with \GH's API \emph{directly} by sending data representing comments \emph{only} from issues matching a particular query string. Moreover, \tool\ cleans, preprocesses, and tokenizes the \emph{automatically} retrieved \GH\ data and subsequently runs the sanitized issue comment threads through \citeauthor{Arya2019}'s model for classification. Further, the model
may be interchanged
(q.v.~\cref{sec:nlp}).

\citet{Karantonis2021} provides a multi-label prediction for \GH\ issues using the \citetitle{Liu2019} NLP model~\cite{Liu2019}. Their approach, however, is for automatically assigning issue \emph{labels} (e.g., ``bug,'' ``feature''), whereas ours classifies issue \emph{comments} based on a keyword-based query string.
Also,
unlike \citeauthor{Karantonis2021}, who strictly relies on a Python notebook interface, \tool\ provides an (optionally interactive) CLI UI using the \ti{PyInquirer} library, enabling an interactive command menu and command-line arguments. Furthermore, \tool\ automatically authenticates with \GH's API using a supplied access token and checks for the remaining API query limit.

\citet{Fadhel2018} writes a blog post and associated iPython notebook~\cite{Fadhel2020} describing how to classifying discussions within code reviews that are part of \GH\ pull requests. Similar to \citet{Karantonis2021}, there is no \GH\ integration---users must enter the data manually---and no query feature. Although pull requests are treated similarly to issues in \GH, \citeauthor{Fadhel2018}'s classification model is highly-tuned to code review discussions, which my not be entirely amenable to our stated use case of studying, e.g., usage of particular APIs.

\section{Conclusion \& Future Work}\label{sec:conc}

An open-source, publicly available tool~\cite{Paing2021}---as part of a broader research infrastructure---to help ESE researchers quantify the types of discussions in \GH\ issue comment threads around a particular query string of interest has been demonstrated. \tool\ is implemented in Python and uses libraries, e.g., \ti{NLTK}, for string preprocessing and NLP model loading to automatically classify each of the strings. \tool\ also interfaces with \GH's API and features a (optionally interactive) CLI UI\@. As the tool is in its early stages, plans for a fuller evaluation were discussed.

In the future, we plan to expand onto other platforms such as Stack Overflow to also process developer Q\&A posts.
To further enhance performance, we will classify each issue thread as they are retrieved from \GH's API instead of waiting for all to be retrieved.
Other future plans include exploring alternate tool forms, e.g., browser extensions, and performing a thorough evaluation.

\printbibliography%

\end{document}